\documentstyle[12pt]{article}
\begin{document}

\def\be{\begin{equation}}
\def\ee{\end{equation}}
\def\bea{\begin{eqnarray}}
\def\eea{\end{eqnarray}}
\def\fmn{{\cal F}_{\mu\nu}}
\def\zz{{\cal Z}[J]}
\def\am{{\cal A}_{\mu}}
\def\ama{\am^{a}}
\def\hact{\Bigl[\frac{\delta S_0}{\delta {\cal A}_{\rho}^a(x)}\Bigr]^2}
\def\gda{\; det{\cal M}({\cal A})}
\def\gdam{\; det^{-1}{\cal M}({\cal A})}
\def\dd{\prod_{x}}
\def\pdm{\partial_{\mu}} \def\nm{\nabla_{\mu}}
\def\soa{\frac{\delta^2 S_0}{\delta{\cal A}^a_\rho (x)\delta \am^b (y)}}
\def\cj{c_{\jmath}}
\title{
\bf \mbox{} \\ A simplified version of Higher Covariant Derivative
regularization.} \vspace{.5cm}
\author{{\bf T.D.Bakeyev}
\date{}
\\ {\it Moscow State University} }\maketitle

\begin{abstract}

A simplified version of Higher Covariant Derivative regularization
for Yang-Mills theory is constructed. This may make Higher Covariant
Derivative method more attractive for practical calculations.
\end{abstract}

\section{Introduction.}

The construction of invariant regularizations for gauge theories is very
important both for the practical calculations and for the investigation of
interesting nonperturbative phenomena. The most popular scheme as yet
is a dimensional regularization \cite{Dim}. The drawback of this regularization
is an absence of obvious generalization to nonperturbative approaches and
inapplicability to chiral and supersymmetric models. The lattice formulation
\cite{Wils} preserves gauge invariance and has nonperturbative meaning, but
it also does not seem to be applicable for the regularization of chiral,
topological and supersymmetric theories.

The regularization of gauge theories by higher covariant derivatives
\cite{Sl1,Zinn} and gauge invariant Pauli-Villars regulators \cite{Sl3} is
an alternative nonperturbative regularization scheme which has an advantage
of being applicable to chiral and supersymmetric models. In Ref.\cite{BakSl}
we constructed an unambiguous regularized functional which avoided the
objections raised in Refs. \cite{Ruiz,Warr,Sen}.

In this paper we obtain a more compact and convenient expression for the
regularized Lagrangian which is simpler and requires much less computational
efforts than the one in Ref.\cite{BakSl}. This may make the Higher
Covariant Derivative method more attractive for the practical calculations
in models where dimensional regularization is not applicable.

\section{Obtaining the regularized functional.}

One of the ways to regularize the theory is to modify the propagators
by introducing into Lagrangian higher derivative terms. However this procedure
breaks gauge invariance. To preserve the symmetry one can add into
the Yang-Mills(YM) Lagrangian the terms containing
higher covariant derivatives \cite{Sl1,Zinn}.

We choose the regularized Lagrangian in the form proposed in \cite{BakSl}:
\be
{\cal L}_{YM}(x) \rightarrow {\cal L}^{\Lambda}(x)=
{\cal L}_{YM}(x)+\frac{1}{4\Lambda^{10}} \hact
\label{1}\ee
where:
\be {\cal L}_{YM}=\frac{1}{8}{\bf tr}\fmn^2 \;\;\; ;\;\;\;
S_0=\int (\nabla^2 \fmn^a )^2 dx \label{3}\ee
Here $\fmn$ is the usual curvature tensor and $\nabla$ is the
covariant derivative:
\be \fmn^a=\partial_\nu\ama-\pdm {\cal A}_\nu^a+t^{abc}\am^b{\cal A}_\nu^c
\label{3a}\ee
\be \nm^{ab}=\pdm\delta^{ab}+t^{abc}\am^c\label{3b}\ee

Then the generating functional for the case of the Lorentz gauge
can be written as follows:

\bea &&\zz=\int\exp\Bigl\{\imath\int\Bigl[{\cal L}_{YM}(x)
+J_{\mu}^a(x)\ama(x) \Bigr]dx+\nonumber\\&&+
\frac{\imath}{4\Lambda^{10}}\int \hact dx\Bigr\}
\delta(\pdm\am)\gda\dd D\am\label{4}\eea
where ${\cal M}^{ab}=\pdm\nm^{ab}$.
It is more convenient for us to transform this functional to another form.
Firstly along the lines of Ref.\cite{BakSl} we bring in the integration
over the auxiliary fields $h_\rho$:

\bea &&\zz=\int\exp\Bigl\{\imath\int\Bigl[{\cal L}_{YM}(x)+
J_{\mu}^a(x)\ama(x)-\frac{1}{2}(h_\rho^a(x))^2\Bigr]dx+\nonumber\\&&+
\frac{\imath}{2\Lambda^5}\int h_{\rho}^a(x)
\frac{\delta S_0}{\delta {\cal A}_\rho^a(x)} dx\Bigr\}
\delta (\pdm\am)\gda\dd Dh_\rho D\am \label{6}\eea
To separate the integration over covariant transversal and longitudinal parts
of $h_\rho$ we multiply the functional (\ref{6}) by "unity":

\be det \nabla^2 \int\dd \delta (\nm(h_\mu+\nm u))Du(x)=1
\label{7}\ee
and change variables:
\be h_\mu\rightarrow h_\mu-\nm u \label{8}\ee
Then we get:

\bea &&\zz=\int\exp\Bigl\{\imath\int\Bigl[{\cal L}_{YM}(x)+
J_{\mu}^a(x)\ama(x)-\frac{1}{2}
(h_{\rho}^a-\nabla_\rho u^a)_x^2\Bigr]dx+\nonumber\\&&+
\frac{\imath}{2\Lambda^5}\int (h_{\rho}^a-\nabla_\rho u^a)_x
\frac{\delta S_0}{\delta {\cal A}_{\rho}^a(x)} dx\Bigr\}
\times\nonumber\\&&\times
\delta (\pdm\am)\delta(\nm h_\mu)\gda det \nabla^2 \dd Du Dh_\rho D\am
\label{9}\eea
Due to the gauge invariance of $S_0$:
\be \int \nabla_\rho u^a(x)\frac{\delta S_0}{\delta {\cal A}_{\rho}^a(x)} dx=0
\label{10}\ee
only covariant transversal part of $h_\rho$ gives nontrivial contribution to
eq.(\ref{6}). On the surface $\nm h_\mu=0$ we have:
\be \int h_\rho^a(x)\nabla_\rho u^a(x)dx=0
\label{11}\ee
Using Eqs. (\ref{10},\ref{11}) we can omit these terms from the exponent
of equation (\ref{9}). Integrating over $u(x)$ we get:

\bea &&\zz=\int\exp\Bigl\{\imath\int\Bigl[{\cal L}_{YM}(x)+
J_{\mu}^a(x)\ama(x)-\frac{1}{2}(h_{\rho}^a(x))^2+
\frac{1}{2\Lambda^5}\int h_{\rho}^a(x)
\frac{\delta S_0}{\delta {\cal A}_\rho^a(x)}\Bigr] dx\Bigr\}
\times\nonumber\\&&\times
\delta(\nm h_\mu)\delta(\partial_\rho {\cal A}_\rho)
\gda det^{\frac{1}{2}} \nabla^2 \dd Dh_\mu D {\cal A}_\rho
\label{12}\eea
The free propagators generated by the exponent in Eq.(\ref{12})
have the following UV behaviour: $\widehat{\am {\cal A}_\nu}\sim k^{-12}$;
$\widehat{h_{\rho}h_\sigma}\sim k^{-10}$; $\widehat{\am h_\rho}\sim k^{-6}$.
Using the functional (\ref{12}) one can see that the divergency
index of arbitrary diagram is equal to:

\be \omega_1=4-4(I-1)-4N_{\cal A}-6L_{\cal A}-4L_h-E_{\cal A}-E_h-3E_{gh}
-\frac{7}{2}E_\delta
\label{5a}\ee
where $I$ is the number of loops, $L_{\cal A}$,$L_h$ are the numbers
of internal and $E_{\cal A}$,$E_h$ are the numbers of external lines
of transversal parts of the fields ${\cal A}$ and $h$ correspondingly,
$N_{\cal A}$ is the number of vertices generated by ${\cal L}_{YM}$,
$E_{gh}$ is the number of external lines of the fields representing $\gda$
and $det^{\frac{1}{2}} \nabla^2$,
$E_\delta=E_h^{long}+E_\lambda$ is the number of external lines representing
longitudinal part of field $h$ and auxiliary field $\lambda$ arising in
the representation of delta-function:
\be\delta(\nm h_\mu)=\int\exp\Bigl\{\int\lambda\nm h_\mu dx\Bigr\}\dd\lambda
\label{add1}\ee
($E_{gh}$ and $E_\delta$ are even numbers). One can see that if $E_h>0$,
the diagram is convergent as it includes at least one internal line of
${\cal A}$ field.
The only divergent graphs are the one loop diagrams with
$L_{\cal A}=L_h=N_{\cal A}=0$, $E_h=E_{gh}=E_\delta=0$ and $E_{\cal A}=2,3,4$.
Therefore the sum of all one-loop divergent diagrams of the functional
(\ref{12}) with external gauge field lines ${\cal A}$ can be presented as
follows:
\bea &&Z_{div}[{\cal A}]=\int\exp\Bigl\{\frac{\imath}{2\Lambda^5}\int\!\!\int
h_\rho^a(x)\soa q_\mu^b(y)dxdy+\dots\Bigr\}\delta(\nm h_\mu)
\delta(\partial_\rho q_\rho)\nonumber\\&&\dd Dh_\mu Dq_\mu\gda
det^{\frac{1}{2}}\nabla^2
\label{28}\eea
Here $\dots$ denotes the terms which provide the infrared convergence
of the integral (\ref{28}) and have no influence on its ultraviolet behavior.

Although higher loop diagrams acquire by power counting a negative
superficial divergent dimension, the divergencies of one loop diagrams are
not smoothed and these diagrams require some additional regularization.

It was proposed in the paper \cite{Sl3} (see also \cite{SlF}) that such
a regularization may be provided by a modified Pauli-Villars(PV) procedure.

In this article we introduce the PV interaction which satisfies the following
conditions:

{\bf A)}it is gauge invariant;

{\bf B)}it completely decouples from the physical fields
in the limit when the masses of PV fields go to infinity;

{\bf C)}it exactly compensates the remaining divergencies of the functional
(\ref{12}) and the cancelation of divergent contributions holds for individual
diagrams
(i.e. the divergent diagrams of the functional (\ref{28}) and PV interaction
have the same structure and one needs no auxiliary regulator).

{\bf D)}when integrated over the fields $\am$ it does not produce any new
divergent subgraphs in the multiloop diagrams (this is known as the problem
of overlapping divergencies, see for example \cite{SlF,AsF2}).

The PV functional with such properties and external gauge field lines
${\cal A}$ looks as follows:

\bea
&&I_{PV}({\cal A})=\int\exp\Bigl\{\frac{\imath}{2\Lambda^5}\int\!\!\int
\overline{B}_\rho^a(x)\soa (B_\mu-\nm\nabla^{-2}\nabla_\nu B_\nu)^{b}_{y} dxdy
+\nonumber\\&&
+\imath M\int \overline{B}_\mu^a(x)B_\mu^a(x)dx\Bigr\}
\delta(\nm\overline{B}_\mu)\delta(\partial_\rho B_\rho)
\dd D\overline{B}_\mu DB_\rho \gdam\nonumber\\&& \prod_{\jmath}
det^{\frac{\cj}{2}} (\nabla^2-M_\jmath^2) \label{13}\eea
Here $\overline{B}_\mu$, $B_\mu$ are anticommuting PV fields, and
PV conditions hold:

\be \sum_{\jmath} \cj =-1 \;\;\; ;\;\;\;
\sum_{\jmath} \cj M_{\jmath}^2=0 \label{15} \ee

Let us check that the functional (\ref{13}) satisfies
all the four conditions imposed above.

{\bf A)} To demonstrate the gauge invariance of the functional (\ref{13})
let us make the change of variables:

\be B_\mu=B_\mu^{tr}+\nm\psi \label{16}\ee
where:
\be B_\mu^{tr}=B_\mu-\nm\nabla^{-2}\nabla_\nu B_\nu \; ;\;\;\;
\psi=\nabla^{-2}\nabla_\nu B_\nu \; ;\;\;\;\nm B_\mu^{tr}=0
\label{17}\ee
and $B_\mu^{tr}$,$\psi$ are anticommuting fields. The Jacobian
of the transformation (\ref{16}) is equal to
$det^{-\frac{1}{2}}\nabla^2$ (the inverse sign of the power of determinant is
due to the Grassmannian nature of variables).
Then the functional (\ref{13}) acquires the form:

\bea
&&I_{PV}({\cal A})=\int\exp\Bigl\{\frac{\imath}{2\Lambda^5}\int\!\!\int
\overline{B}_\rho^a(x)\soa (B_\mu^{tr})^{b}_{y} dxdy +
\nonumber\\&&
+\imath M\int \overline{B}_\mu^a(x)(B_\mu^{tr}+\nm\psi)^a_x dx\Bigr\}
\delta(\nm\overline{B}_\mu)
\delta(\partial_\rho B^{tr}_\rho+{\cal M}\psi)
\dd D\overline{B}_\mu DB^{tr}_\mu D\psi \times\nonumber\\&&\times
\gdam det^{-\frac{1}{2}}\nabla^2 \prod_{\jmath}
det^{\frac{\cj}{2}}(\nabla^2-M_\jmath^2)
\label{18}\eea
One sees that the only $\psi$-dependent term in the exponent of the
expression (\ref{18}) is equal to zero on the surface
$\nm\overline{B}_\mu=0$:
\be \int\overline{B}_\mu^a(x)\nm\psi^a(x)dx=0 \label{19}\ee
This allows us to rewrite the functional (\ref{18}) in the form:

\be I_{PV}({\cal A})=\int I^1({\cal A},B^{tr})I^2({\cal A},B^{tr})\dd DB^{tr}
\label{20}\ee
where:

\bea
&&I^1({\cal A},B^{tr})=\int\exp\Bigl\{\frac{\imath}{2\Lambda^5}\int\!\!\int
\overline{B}_\rho^a(x)\soa(B_\mu^{tr})^b_y dxdy+\nonumber\\&&
+\imath M\int \overline{B}_\mu^a(x)(B_\mu^{tr})^a_x dx\Bigr\}
\delta(\nm\overline{B}_\mu)
\dd D\overline{B}_\mu det^{-\frac{1}{2}}\nabla^2
\prod_{\jmath}det^{\frac{\cj}{2}}(\nabla^2-M_\jmath^2)
\label{21}\eea
and
\be I_2({\cal A},B^{tr})=
\gdam\int\delta(\partial_\rho B^{tr}_\rho+{\cal M}\psi)\dd D\psi
\label{22}\ee

The functional $I^1({\cal A},B^{tr})$ is invariant under simultaneous
transformations:

\be \left\{ \begin{array}{lll}
    \am\rightarrow\am +[\am,\epsilon] +\pdm \epsilon \\
    \overline{B}_\mu\rightarrow\overline{B}_\mu+[\overline{B}_\mu,\epsilon] \\
    B^{tr}_\mu\rightarrow B^{tr}_\mu+[B^{tr}_\mu,\epsilon]
    \end{array}
    \right. \label{23} \ee
The functional $I^2({\cal A},B^{tr})$ is also gauge invariant because
after the integration over $\psi$ it is equal to nonessential constant
and actually it does not depend on ${\cal A}$ and $B^{tr}$ (remember that
$\psi$ is Grassmannian field and integration over $\psi$ in (\ref{22})
produces $\gda$).
That proves the gauge invariance of the functional (\ref{13}).

The demonsration of the gauge invariance of the PV interaction (\ref{13})
is the most essential point of this paper. It allows us to avoid a rather
complicated two-step procedure which we used in Ref.\cite{BakSl} and to make
much simpler the resulting expression for the regularized functional.

{\bf B)}The basic requirement for any regularization is that the terms
of the effective action which in a certain order of perturbation theory
are finite in the original theory, should recover their exact finite values
after the removal of regulating mass parameters (see Ref.\cite{AsF}).
In particular, the regularized partition function must converge to the
original one on a formal level. Here we are going to demonstrate that in
the limit $M\rightarrow\infty$, $M_\jmath\rightarrow\infty$ the PV fields
of the functional (\ref{13}) decouple from the physical fields and contribute
only to local counterterms. Indeed, rescaling the fields in the expression
(\ref{13})
\be \overline{B}_\mu\rightarrow\frac{1}{\sqrt{M}}\overline{B}_\mu
\;\; ;\;\;\; B_\mu\rightarrow\frac{1}{\sqrt{M}} B_\mu
\label{24}\ee
we get:

\bea &&\lim_{M\rightarrow\infty ;\; M_\jmath\rightarrow\infty}I_{PV}({\cal A})=
\int\exp\Bigl\{\imath\int\overline{B}_\mu^a(x)B_\mu^a(x)dx\Bigr\}
\delta(\nm\overline{B}_\mu)\delta(\partial_\rho B_\rho)
\dd D\overline{B}_\mu DB_\mu \times\nonumber\\&&\times
\gdam \label{25}\eea
Making again the change of variables (\ref{16}) for both fields
$\overline{B}_\mu$,$B_\mu$ (the Jacobian of this transformation is equal to
$det^{-1}\nabla^2$) we see that:

\bea &&\lim_{M\rightarrow\infty ;\; M_\jmath\rightarrow\infty}I_{PV}({\cal A})=
\int\exp\Bigl\{\imath\int(\overline{B}^{tr}_\mu+\nm\overline{\psi})^a_x
(B^{tr}_\mu+\nm\psi)^a_x dx\Bigr\} \times\nonumber\\&&\times
\delta(\nabla^2\overline{\psi})\delta(\partial_\rho B^{tr}_\rho+{\cal M}\psi)
\dd D\overline{B}^{tr}_\mu DB^{tr}_\mu D\overline{\psi}D\psi\gdam
det^{-1}\nabla^2=\nonumber\\&& =
\int\exp\Bigl\{\imath\int\overline{B}^{tr}_\mu(x)
B^{tr}_\mu(x) dx\Bigr\} \dd D\overline{B}^{tr}_\mu DB^{tr}_\mu ={\it const}
\label{26}\eea

So the PV interaction (\ref{13}) completely decouples from the physical fields
in the limit $M\rightarrow\infty$,$M_\jmath\rightarrow\infty$.

{\bf C)} Now we shall show that the PV functional (\ref{13}) exactly
compensates the remaining one-loop divergencies of the functional (\ref{12}).
One can see that due to the gauge invariance of $S_0$ the nonlocal
term in the exponent of the expression (\ref{13}) can be rewritten in the form:

\be \int\!\!\int
\overline{B}_\rho^a(x)\soa(\nm\nabla^{-2}\nabla_\nu B_\nu)^b_y dxdy=
\int\overline{B}_\rho^a(x)
\frac{\delta S_0}{\delta {\cal A}_\rho^b(x)} gt^{abc}(\nabla^{-2}
\nabla_\nu B_\nu)^c_x dx \label{27}\ee
and generates only convergent diagrams, as some derivatives in the
r.h.s. of Eq.(\ref{27}) act on external fields $\am$.
Therefore the divergent diagrams generated by Eqs.(\ref{28}) and (\ref{13})
have the same structure (if $\overline{B}_\mu$ corresponds to $h_\mu$
and $B_\rho$ corresponds to $q_\rho$) and their sum is finite.
This fact proves that the PV functional (\ref{13}) compensates
the remaining one-loop divergencies of the functional (\ref{12}).

{\bf D)}At this point we shall show that when integrated over the fields $\am$
the PV functional (\ref{13}) does not produce any new divergent subgraphs
in multiloop diagrams and the generating functional (\ref{12})
regularized by the PV interaction (\ref{13}) is finite
not only at one-loop level, but also for any number of loops.
Let us write this regularized functional, which is the final output
of our paper:

\bea &&\zz=\int\exp\Bigl\{\imath\int\Bigl[{\cal L}_{YM}(x)+
J_{\mu}^a(x)\ama(x)-\frac{1}{2}(h_\rho^a(x))^2+\frac{1}{2\Lambda^5}
h_{\rho}^a(x)\frac{\delta S_0}{\delta {\cal A}_{\rho}^a(x)}\Bigr] dx+
\nonumber\\&&+ \frac{\imath}{2\Lambda^5}\int\!\!\int
\overline{B}_\rho^a(x)\soa(B_\mu-\nm\nabla^{-2}\nabla_\nu B_\nu)^b_y dxdy+
\imath M\int \overline{B}_\mu^a(x)B_\mu^a(x)dx\Bigr\}\times\nonumber\\&&
\times\delta(\nm h_\mu) \delta(\nm\overline{B}_\mu)
\delta(\partial_\rho {\cal A}_\rho) \delta(\partial_\rho B_\rho)
det^{\frac{1}{2}} \nabla^2 \prod_{\jmath}
det^{\frac{\cj}{2}}(\nabla^2-M_\jmath^2)
\dd  D\overline{B}_\mu DB_\rho Dh_\mu D{\cal A}_\rho\nonumber\\&&
\label{29}\eea
One can see that the divergency index of arbitrary diagram generated by
the functional (\ref{29}) is much the same as in expression (\ref{5a})
except for the contribution of external lines of PV fields:

\be \omega_2=\omega_1-E_B-E_{\overline{B}}-3E_{gh}^{PV}
-\frac{7}{2}E_\delta^{PV}
\label{30}\ee
where $E_B=E_{\overline{B}}$ are the numbers of external lines of
transversal parts of the fields $B$ and $\overline{B}$ correspondingly,
$E_{gh}^{PV}$ is the number of external lines of the fields representing
\newline
$det^{\frac{\cj}{2}}(\nabla^2-M_\jmath^2)$,
$E_\delta^{PV}=E_{\overline{B}}^{long}+E_\lambda^{PV}$ is the number of
external lines representing longitudinal part of field
$\overline{B}$ and auxiliary field $\lambda^{PV}$ arising in
the representation of delta-function:
\be\delta(\nm\overline{B}_\mu)=\int\exp\Bigl\{\int\lambda^{PV}
\nm\overline{B}_\mu dx\Bigr\}\dd\lambda^{PV}
\label{add2}\ee
($E_{gh}^{PV}$ and $E_\delta^{PV}$ are even numbers). One can see that
if $E_B>0$, the diagram is convergent, as it includes at least one
internal line of ${\cal A}$ field.
So for the diagram to be divergent the following conditions must hold:
$E_B=E_{\overline{B}}=0$, $E_{gh}^{PV}=E_\delta^{PV}=0$ and the number of loops
$I=1$. As was shown above, all divergencies in such one-loop diagrams
generated by the functional (\ref{29}) are compensated becouse of the
application of PV procedure.

The absence of new divergent diagrams generated by the PV interaction when
integrated over $\am$ is due to the fact that the propagators of
the fields $\am$ decrease for large momenta faster than the propagators
of PV fields. Such a solution of the problem of overlapping divergencies
was proposed in Ref.\cite{SlF} and realized in Ref.\cite{BakSl}.
This way of solving the problem imposes certain restrictions on the form
of higher covariant derivative term in the intermediate Lagrangian (\ref{1}).
As was pointed out in Ref.\cite{BakSl}, if one is interested in calculation of
only one-loop diagrams, one can use a much more simple intermediate Lagrangian:

\be
{\cal L}_{YM}(x) \rightarrow {\cal L}^{\Lambda}(x)=
{\cal L}_{YM}(x)+\frac{1}{8\Lambda^{4}}{\bf tr}(\nabla^2 \fmn)^2
\label{31}\ee
(Qua the examples of such calculations see Refs.\cite{Ruiz,Pronin}.)
However starting from Lagrangian (\ref{31}), one can not introduce
the "correct" PV interaction which does not generate additional divergent
subgraphs in multiloop diagrams (overlapping divergencies) when
integrated over $\am$ (for details see Ref.\cite{BakSl}).
So if one is interested in generating functional which is finite in any
order of perturbation theory and also has nonperturbative meaning,
one should use the functional (\ref{29}).

Let us make a conclusion to this section. We constructed regularized
generating functional (\ref{29}) which is "correct" in a sense that:

{\bf A)} It satisfies the correct Slavnov-Taylor identities
(see Refs.\cite{Sl5,Tail}), as we used only gauge invariant regularizing terms.

{\bf B)}In the limit $M\rightarrow\infty\; ;\;M_\jmath\rightarrow\infty$
and then $\Lambda\rightarrow\infty$ all unphysical excitations decouple
from the physical fields and contribute only to local counterterms.

{\bf C and D)}All the diagrams generated by the functional (\ref{29})
are finite as the integal over anticommuting fields $\overline{B}$,$B$
subtracts the divergent one-loop diagrams which arise due to integration over
${\cal A}$,$h$ and no divergent subgraphs in multiloop diagrams
(overlapping divergencies) are present.

At the same time the functional (\ref{29}) is simpler that the
one obtained in Ref.\cite{BakSl} and requires much less
computational effort, so it can be more attractive for using in practical
calculations.

Finally we notice that the nonlocal term (\ref{27}) in the expression
(\ref{29}) does not contribute in the limit $M\rightarrow\infty$.
For finite $M$ we have shown it produces finite diagrams, and in the limit
$M\rightarrow\infty$ its contribution disappears. Being interested
finally in the limit $M\rightarrow\infty$ we can omit this term
in the Eq.(\ref{29}). It simplifies the final expression for the regularized
functional and make the effective action local. Omitting this term we break
the gauge invariance for finite $M$. The Slavnov-Taylor identities will be
violated by finite terms of order $O(M^{-1})$. These terms are harmless
as they have no influence on the counterterms and disappear in the limit
$M\rightarrow\infty$.

\section{Discussion.}

In this paper we constructed a consistent invariant regularization of
gauge models withing the framework of Higher Covariant Derivative method.
The obtained formulation is considerably simpler and requires much
less computational effort than the one in Ref.\cite{BakSl}.
This may make the regularization of gauge theories by higher covariant
derivatives and gauge invariant Pauli-Villars regulators much more attractive
for using in practical calculations in models where dimensional regularization
is not applicable.

\section{Acknowledgments.}

I am very grateful to Prof. Slavnov A.A. (Steklov Mathematical Institute)
for generous help and numerous discussions.
This research was supported by Russian Basic Research Fund
under grant 96-01-00551 and Presidential grant for support of
leading scientific schools.

  \end{document}